\documentclass{raa}            % referee version: for submission

\usepackage{graphicx,times}             %for PS/EPS graphics inclusion, new

\usepackage{natbib}
\usepackage{amssymb,amsmath}
\bibpunct{(}{)}{;}{a}{}{,}

\usepackage[pagebackref=true]{hyperref}

\usepackage{lineno}

\begin{document}

  \title{Measurements of the solar coronal magnetic field based on coronal seismology with propagating Alfv\'{e}nic waves: forward modeling}
%   \subtitle{I. Place Your Subtitle Here}

   \volnopage{Vol.0 (20xx) No.0, 000--000}      %%preserved for Editor. DOn't remove!
   \setcounter{page}{1}          %%starting page, preserved for Editor. DOn't remove!

   \author{Yuhang Gao %% Put your Chinese name in "( )" if you like. Note to open line 11 "\usepackage[UTF8]{ctex}"
      \inst{1,2}
   \and Hui Tian
      \inst{1}
   \and Tom Van Doorsselaere
      \inst{2}
    \and Zihao Yang
        \inst{1,3}
    \and Mingzhe Guo
        \inst{2,4}
    \and Konstantinos Karampelas
        \inst{2}
   }
%% Here is an example of three authors come from different institutes.
%% For single author or all the authors from an institute, use "\inst{}" only

   \institute{School of Earth and Space Sciences, Peking University, Beijing, 100871, People's Republic of China; {\it huitian@pku.edu.cn}\\
%% Please give the E-mail address of the author, to whom future correspondence and
%% offprint requests will be sent.
        \and
             Centre for mathematical Plasma Astrophysics, Department of Mathematics, KU Leuven, Celestijnenlaan 200B bus 2400, B-3001 Leuven, Belgium\\
        \and
            High Altitude Observatory, National Center for Atmospheric Research, Boulder, CO 80307, USA\\
        \and
             Shandong Provincial Key Laboratory of Optical Astronomy and Solar-Terrestrial Environment, Institute of Space Sciences, Shandong University, Weihai 264209, China\\
\vs\no
   {\small Received 20xx month day; accepted 20xx month day}}

\abstract{Recent observations have demonstrated the capability of mapping the solar coronal magnetic field using the technique of coronal seismology based on the ubiquitous propagating Alfv\'{e}nic/kink waves through imaging spectroscopy. We established a magnetohydrodynamic (MHD) model of a gravitationally stratified open magnetic flux tube, exciting kink waves propagating upwards along the tube. Forward modeling was performed to synthesize the Fe \textsc{xiii} 1074.7 and 1079.8 nm spectral line profiles, which were then used to determine the wave phase speed, plasma density, and magnetic field with seismology method. A comparison between the seismologically inferred results and the corresponding input values verifies the reliability of the seismology method. In addition, we also identified some factors that could lead to errors during magnetic field measurements. Our results may serve as a valuable reference for current and future coronal magnetic field measurements based on observations of propagating kink waves.
\keywords{Sun: corona --- Sun: magnetic fields --- magnetohydrodynamics}
}

   \authorrunning{Y. Gao et al. }            %author_head in even pages
   \titlerunning{Coronal seismology with propagating Alfv\'{e}nic waves}  % title_head in odd pages

   \maketitle
%% The author head (on even pages) and the title head (on odd pages) will be
%% automatically extracted from \author{} and \title{}. Whenever the title is too long,
%% you will be asked to supply a shorter one by inserting either \authorrunning{} or
%% \titlerunning{} before \maketitle. Anyway, you can specify your own heads.
%%
%%
%% Note: In the following text body of your manuscript, please note several differences from
%%       other major journals:
%% (1) \subsection{Please Capitalize the First Letter of Each Notional Word in Subsection Title}
%% (2) Please Capitalize the First Letter of Each Notional Word in all tables' captions

%
%________________________________________________ sections below
%
\section{Introduction}           %% first-level sections will be auto-capitalized
\label{sect:intro}

The magnetic field plays a crucial role in various physical processes in the solar and stellar coronae. The dissipation of magnetic energy is believed to drive solar eruptive events (e.g., flares and coronal mass ejections) and cause heating of the corona. While the magnetic field in the lower solar atmosphere can be reliably measured through spectro-polarimetric observations, direct measurements of the coronal magnetic field have remained challenging for decades. Several methods, including the spectro-polarimetry of coronal infrared lines \citep{lin2000,lin2004,schad2024}, coronal radio observations \citep[e.g.,][]{Fleishman2020,Chen2020,tan2022}, and magnetic-field-induced transitions of extreme ultraviolet emission lines \citep[e.g.,][]{li2015,li2016,landi2020,ChenY2021,chen2023RAA}, have been proposed and attempts of measurements have been made. However, these approaches all face limitations and none of them could be used for routine measurements of the global coronal magnetic field.

Another technique that could be used to measure the coronal magnetic field is coronal seismology. This technique combines the magneto-hydrodynamic (MHD) wave theory with observed wave parameters (e.g., period, amplitude, propagation speed, damping time) to diagnose various physical properties, particularly the magnetic field. Different wave phenomena in the corona have been used for coronal seismology, including standing kink waves in magnetic loops \citep[e.g.][to name but a few]{naka2001,aschwanden2002,tvd2007,tian2012,zhang2022,gao2022,gao2024,zhong2023NatSR,LiDong2023}, propagating slow magneto-acoustic waves \citep[e.g.,][]{jess2016}, sausage waves (\citealp{ChenSX2015,Guo2016}; see \citealp{LiBo2020} for a review), propagating kink waves in streamers \citep{chen2011,Guo2022}, and torsional oscillations in solar surges \citep{kohutova2020}. However, these studies provided only one-dimensional (1D) distributions or single values of the magnetic field in specific coronal structures, such as oscillating loops or streamers. To create global 2D magnetic field maps, we need to utilize ubiquitous and continuous wave phenomena. The pervasive propagating disturbances in Dopplergrams \citep{tomczyk2007,tomczyk2009,liu2015,morton2015,morton2019} observed by the Coronal Multi-channel Polarimeter \citep[CoMP;][]{tomczyk2008} are ideal for this purpose. These propagating disturbances are interpreted as kink or Alfv\'{e}nic waves \citep{tvd2008},
% \footnote{Initially, \cite{tomczyk2007} identified these CoMP waves as Alfvén waves. However, the interpretation of the wave mode soon became under debate. Nowadays, most people believe these waves are actually kink waves, a type of magneto-acoustic waves with the azimuthal wavenumber $m=1$. Despite this, kink waves share similar properties with Alfv\'{e}n waves, as both are transverse MHD waves with a (nearly) incompressible nature and rely on magnetic tension as the primary restoring force \citep[e.g., see][]{tvd2008,goossens2009}. For this reason, these kink waves are also called ``Alfv\'{e}nic waves".}
and their propagation speeds are naturally linked to the local magnetic field. Based on these CoMP observations, \cite{yang2020sci,yang2020ScChE} have successfully measured the global distribution of the coronal magnetic field for the first time.
Following these successful attempts, a routine (continuous) measurement of the global coronal magnetic field based on similar observations from the Upgraded CoMP (UCoMP; \citealp{Landi2016}) has been recently achieved, which allows for the construction of coronal synoptic magnetograms (Carrington maps) \citep{yang2024}. 

The CoMP and UCoMP instruments can conduct spectroscopic observations of the Fe \textsc{xiii} lines at 1074.7 and 1079.8 nm in the coronal region above the solar limb. From the Dopplergrams of Fe \textsc{xiii} 1074.7 nm, propagating kink waves can be identified throughout the corona. The propagating or phase speed $c_\text{k}$ of these waves (also named the kink speed) could be obtained by constructing a time-distance map of Doppler velocity, while the coronal density can be inferred from the observed Fe \textsc{xiii} 1079.8-nm/1074.7-nm intensity ratio. 

For kink waves, we have 
\begin{equation}\label{eq1}
    c^2_\text{k}=\frac{B_\text{i}^2+B_\text{e}^2}{\mu_0(\rho_\text{i}+\rho_\text{e})}\,,
\end{equation}
where $\mu_0$ is the magnetic permeability, $B$ and $\rho$ are the magnetic field and mass density, respectively. The subscripts i and e refer to parameters inside and outside the magnetic flux tubes (the waveguides). Since CoMP and UCoMP likely cannot resolve individual flux tubes, we can only work with an average density $\langle\rho\rangle$. Furthermore, in the low-$\beta$ coronal environment, the internal and external magnetic fields are often assumed to be approximately equal \citep[see e.g.,][]{tomczyk2009,morton2015,zhong2023NatSR}. This leads to the simplified expression for the kink speed:
\begin{equation}\label{eq2}
    c^2_\text{k}=\frac{B^2}{\mu_0 \langle\rho\rangle}\,,
\end{equation}
which is widely used in estimations of coronal magnetic fields \citep{long2017,yang2020ScChE,yang2020sci,yang2024}.

Given the potential of these measurements to provide routine coronal magnetograms on a daily basis, which could play a crucial role in future solar physics research, it is essential to thoroughly assess the reliability and robustness of the methodology used in \cite{yang2020ScChE,yang2020sci} and \cite{yang2024}. \cite{magyar2018} performed 3D MHD simulations of propagating kink waves under various conditions and conducted forward modeling to evaluate the reliability of this method in deriving the magnetic field strength. Their findings indicated that the magnetic field strengths inferred through seismology closely match the input values, typically with an error less than $\sim$20\%. However, there is a limitation in their simulation. They utilized a non-stratified setup that excluded the effects of gravity, resulting in uniform initial density and propagation speed in the vertical direction.

The gravitational stratification can play a significant role as it causes the Alfv\'{e}n speed and kink speed $c_\text{k}$ to vary with height.
% It also introduces a characteristic length scale in the medium, known as the scale height, which can lead to wave dispersion \citep[see][]{roberts2019mhd}.
% Previous theoretical works \citep[e.g.,][]{dymova2005,Erdelyi2007,lopin2013} have shown that kink waves in a thin flux tube satisfy the following wave equation:
% \begin{equation}
%     \frac{\mathrm{d}v_r^2}{\mathrm{d}z^2}+\frac{\omega^2}{c^2_\text{k}(z)}v_r=0\,,
% \end{equation}
% where $v_r(z)$ is the radial velocity perturbation, and $\omega$ is the circular frequency. 
This variation can affect the wave tracking method which typically relies on a linear fit of velocity signals \citep[e.g., see Figure 6 in][]{tomczyk2009}. Therefore, it is important to assess how the gravitational stratification affects the seismological results and estimate the possible error range. So we conducted 3D MHD simulations of propagating kink waves in stratified coronal open flux tubes. 
Following \cite{magyar2018}, we employed forward modeling to compare the seismologically derived results with the actual values of physical parameters from our simulation. This paper is organized as follows: Section \ref{sec:method} describes our simulation setup and methodology, Section \ref{sec:res} presents the simulation and forward-modeling results, along with detailed comparisons between seismology results and input values, and Section \ref{sec:conclusion} provides a discussion and summary of our findings.

%% Authors can give a citation as 'Michel et al. 1992'.
%% You may also use \cite, \citep and \citet for citation, and use Table~1 or Figure~1
%% and so forth. Using \ref and \label for cross-references of Tables/Figures
%% is a good way in adjusting/adding/removing text, tables or figures.

\section{Method}\label{sec:method}

\begin{figure}
   \centering
   \includegraphics[width=\textwidth, angle=0]{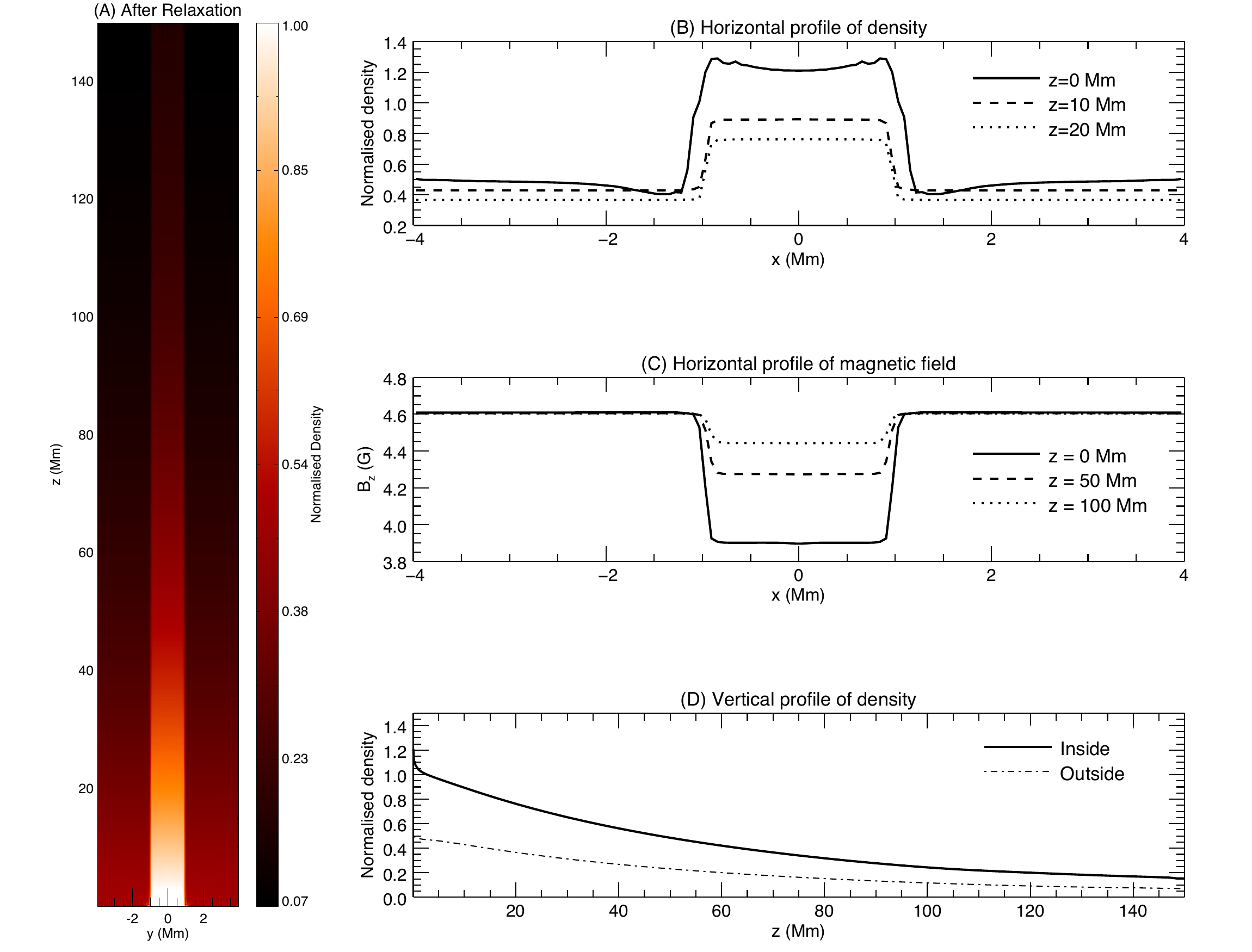}
   \caption{Model of the open magnetic flux tube. (A) Variation at the $x=0$ plane after relaxation. (B)-(C) Horizontal profiles of density and magnetic field along the $x$ axis at three different heights (indicated by different line styles). (D) Vertical profile of density demonstrating gravitational stratification, with solid and dashed lines indicating density inside and outside the flux tube, respectively.}
   \label{Fig1}
\end{figure}

The model used in this study is a gravitationally stratified, open magnetic flux tube with a radius of 1 Mm, similar to that in \cite{gao2024b} (hereafter referred to as Paper I). The main difference is that the initial magnetic field in this study is set at around 4 G, instead of 10 G, to better match the seismologically inferred results in \cite{yang2020sci}. As in Paper I, the magnetic field is oriented in the $z$ direction and remains nearly uniform across the simulation domain, with only small spatial gradients to maintain total pressure balance. As a result, the Alfvén and kink speeds increase with height as the density decreases due to stratification. A relaxation process of 2400 s was conducted to achieve a quasi-magnetohydrostatic state, as shown in Figure \ref{Fig1}. The figure shows that the post-relaxation magnetic field has some spatial variation but mostly within 0.7 G.

The kink wave driver is also similar to that in Paper I, with a velocity amplitude of 8 km s$^{-1}$ and a period of 300 s. The primary velocity perturbation is along the $x$ direction.

We ran the 3D MHD simulation in Cartesian coordinates with the PLUTO code \citep{Mignone2007}. The simulation domain spans [-4, 4] Mm$\times$[-4, 4] Mm$\times$[0, 150] Mm, with a uniform grid of $128\times128\times1024$ cells, providing spatial resolutions of 62.5 km in the horizontal ($x$ and $y$) directions and 146.5 km in the vertical ($z$) direction. We chose a second-order parabolic spatial scheme and a Roe Riemann solver. The boundary conditions were set to be outflow, except for the lower boundary, where the kink wave driver was introduced by adjusting $v_x$ and $v_y$, while other parameters were fixed. As in Paper I, the upper 50 Mm ($z > 100$ Mm) was set as a velocity absorption region (VAR) to minimize numerical reflections from the upper boundary \citep[see also][]{pelouze2023,guo2023,gao2023}. For the subsequent analysis, we only considered the region below $z = 100$ Mm which can give us physical results.

Once the kink wave driver was applied, the propagating waves were rapidly excited, with their properties thoroughly analyzed in Paper I. Here, we focus on assessing the reliability of the seismology method described in Section \ref{sect:intro} using the simulation outputs. To do so, we performed forward calculations to synthesize spectroscopic observables of CoMP and UCoMP, namely, the Fe \textsc{xiii} 1074.7 and 1079.8 nm spectral line profiles. Specifically, we used synthesized observation of the 1074.7 nm line to perform the wave tracking and determine the propagation speed, while synthesized observation of the 1079.8 nm line was only employed for the density diagnostic using the intensity ratio method \citep{yang2020ScChE,yang2020sci}.

We applied the FoMo code \citep{tvd2016} to synthesize the spectral profiles at all pixels in the $yz$ plane. The photo-excitation \citep[see][]{Young2003,yang2020ScChE} was not considered for simplification, as their effects on the spectral lines are not significant at the lower corona. 
The line of sight (LOS) was chosen as the $x$ axis. In this way, we can reconstruct 2D maps (in the $yz$ plane) of the intensity and Doppler velocity by fitting a single Gaussian to each spectral profile. The resulting maps consist of 128 pixels in the $y$ direction from -4 Mm to 4 Mm, and 500 pixels in the $z$ direction from 0 Mm to 100 Mm. Since the primary velocity perturbation is along the $x$ direction (or in the LOS plane), the Doppler velocity maps capture the wave propagation signals, while the intensity maps do not show any transverse displacement. 

\section{Results}\label{sec:res}

\subsection{Before Degrading}\label{subsec:before}

\begin{figure}
   \centering
   \includegraphics[width=\textwidth, angle=0]{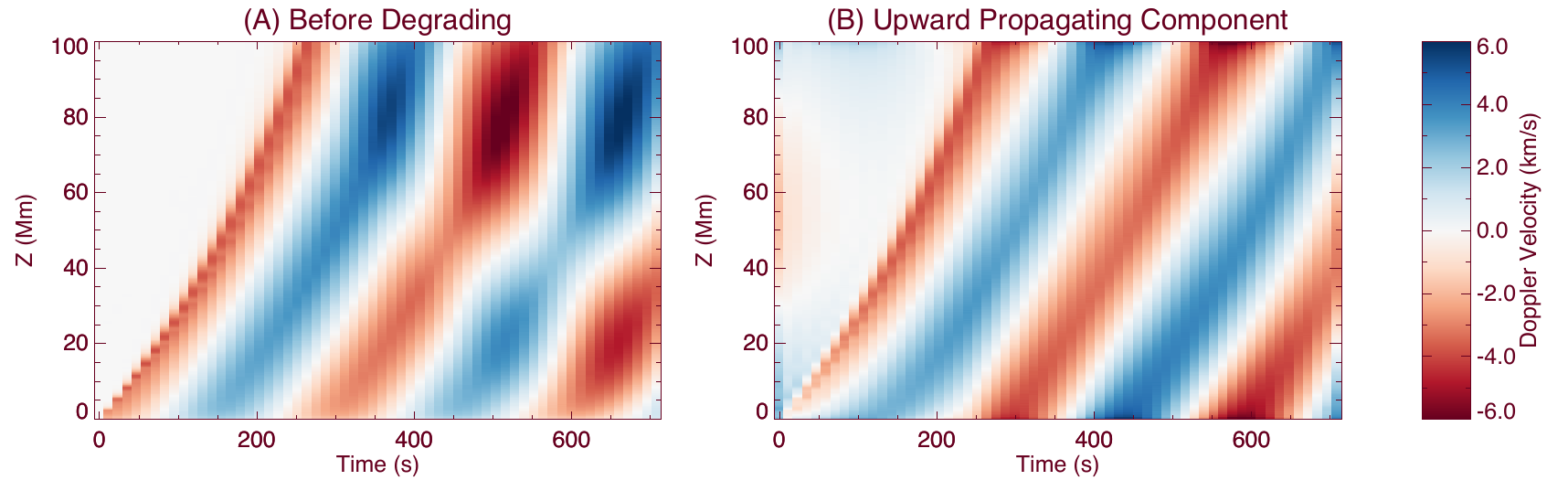}
   \caption{Time distance maps of the Doppler velocity along the $z$ axis at $y=0$. (A) corresponds to the original forward modeling output, while (B) corresponds to the upward propagating component obtained from Fourier filtering.}
   \label{Fig2}
\end{figure}

\begin{figure}
   \centering
   \includegraphics[width=\textwidth, angle=0]{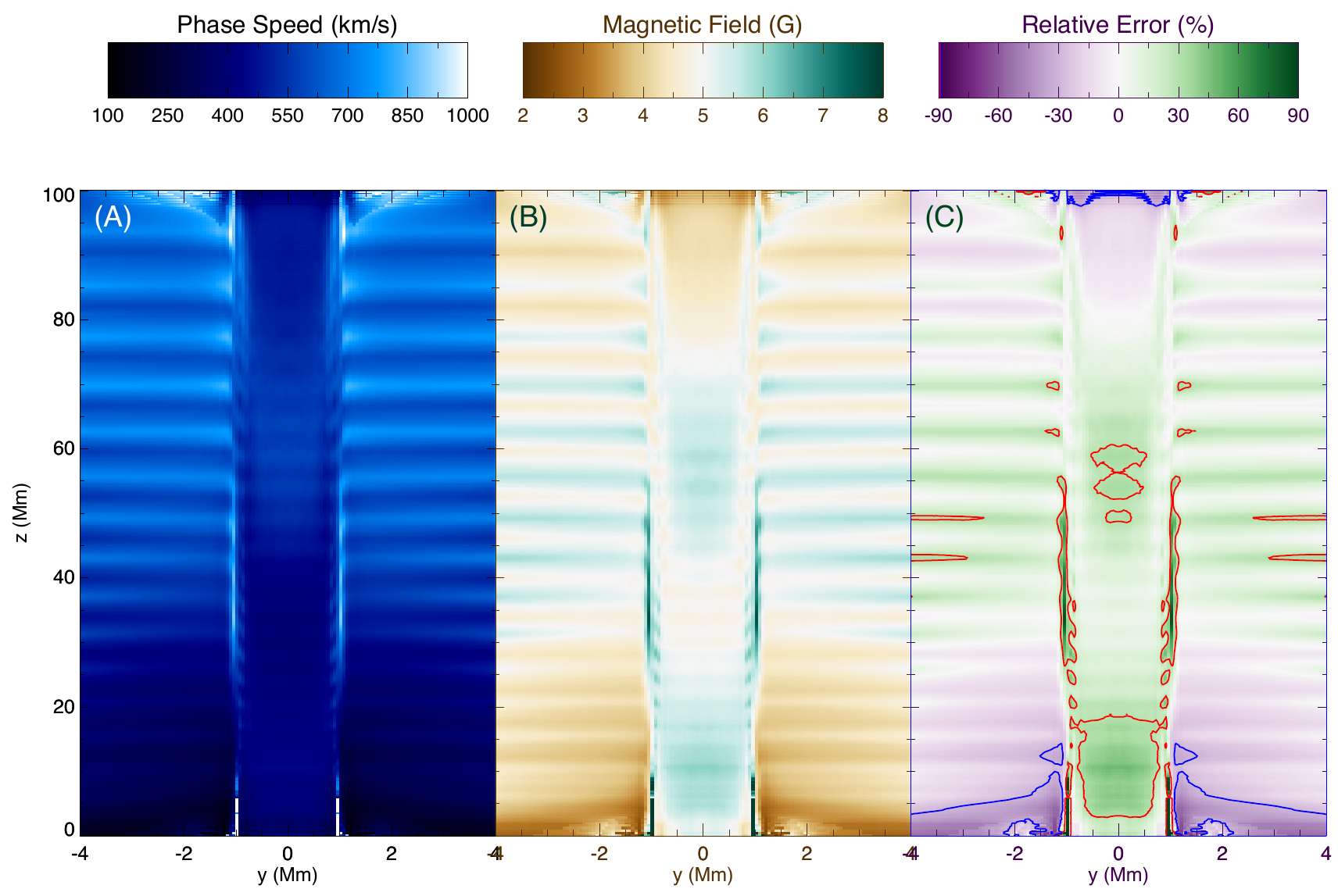}
   \caption{Seismology results for the case with original resolution (before degrading). (A) Spatial distribution of the phase speed. (B) Spatial distribution of the magnetic field ($B_\text{seis}$). (C) Spatial distribution of the relative error of $B_\text{seis}$ compared with the input magnetic field $B_\text{los}$. Blue and red contours correspond to the relative error of -30\% and 30\%, respectively.}
   \label{Fig3}
\end{figure}

\begin{figure}
    \centering
    \includegraphics[width=\linewidth]{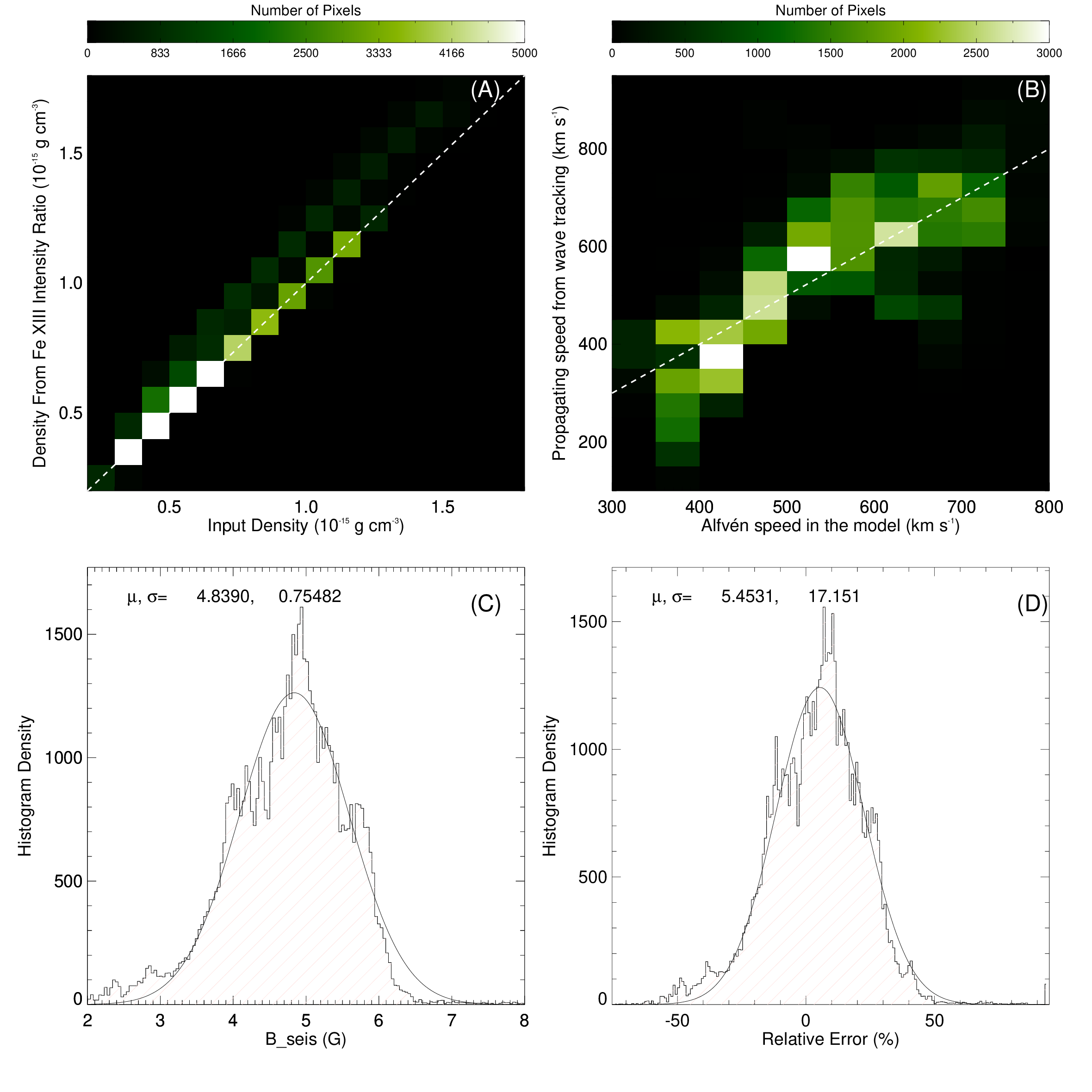}
    \caption{(A) 2D histogram comparing the density derived from the Fe \textsc{xiii} intensity ratio and the input density $\rho_\text{los}$. (B) 2D histogram comparing the propagation speed obtained from wave tracking and the Alfv\'{e}n speed in the model. (C)-(D) Histograms of $B_\text{seis}$ and the relative error. The Gaussian fit results are overplotted, with the mean value ($\mu$) and standard deviation ($\sigma$) indicated at the top.}
    \label{Fig4}
\end{figure}

We first generated time-distance (TD) maps of the Doppler velocity along the $z$ direction, which can be produced at each $y$ position. In Figure \ref{Fig2}(A), the TD map along the flux tube's axis (i.e., $y=0$) is presented, clearly illustrating the propagation of Doppler velocity disturbances. The increasing slope reflects the growing propagation speed with height. However, after 400 s, unusual patterns appear due to wave reflections.  Despite efforts to suppress numerical reflections from the upper boundary at $z=150$ Mm using a VAR (see Section \ref{sec:method}), it is still difficult to eliminate all reflections.
Some reflections may come from the layer of $z=100$ Mm, which is the lower boundary of the VAR; while others could be attributed to the vertical inhomogeneities of density and phase speed. As noted in previous studies, even smooth phase speed gradients can cause partial wave reflection \citep[e.g.,][]{verdini2007,hahn2018,pascoe2022,bose2024}. In fact, real CoMP observations of coronal kink waves often reveal both upward and downward propagating components \citep{tomczyk2009,morton2015}. However, when calculating the propagation speed, these reflected waves can introduce significant errors. To reduce this, we applied the Fourier filtering method to separate the upward and downward propagating wave components \citep[see e.g.,][]{tomczyk2009,threlfall2013,liu2014,tiwari2019,yang2020sci}. Figure \ref{Fig2}(B) shows the TD map for the upward-propagating component, which is used to determine the phase speed.

By applying the wave-tracking method to all pixels (for more details, see e.g., \citealp{yang2020sci}), we obtained a 2D phase speed distribution, as shown in Figure \ref{Fig3}(A). For most pixels, the phase speed falls in the range of 200-800 km s$^{-1}$ (see also Figure \ref{Fig4}(B)), with a general trend of increasing with altitude, though some fluctuations are present (see relevant discussions in Section \ref{sec:conclusion}).

Next, we derived the density by calculating the intensity ratio of the synthesized 1074.7 nm and 1079.8 nm intensity maps. The theoretical relationship between intensity ratio and density can be obtained from the CHIANTI database \citep{dere2023}. The obtained density values were compared with the input values, as shown in Figure \ref{Fig4}(A). 
The input density corresponds to the emissivity-weighted density along the LOS \citep{yang2024}: 
\begin{equation}
    \rho_\text{los}=\frac{\int_{x}\rho(x)\epsilon(x)\mathrm{d}x}{\int_x\epsilon(x)\mathrm{d}x}\,,
\end{equation}
where $\epsilon$ is the Fe XIII 1074.7 nm line emissivity at the corresponding pixel, calculated using the IDL routine \texttt{emiss\_calc.pro} from the CHIANTI software package. The comparison shows that the density derived from the intensity ratio is reliable for most pixels, though about 20\% pixels show an overestimation ($\gtrsim$15\%). However, this overestimation would not lead to large errors in the seismologically inferred magnetic field, as only the square root of density is needed during the calculation.

With the derived phase speed and density, we calculated the magnetic field $B_\text{seis}$. Here we chose to treat the phase speed as the local Alfv\'{e}n speed (i.e., $c_\text{ph}=B_\text{seis}/\sqrt{\mu_0\rho}$), rather than employ Equation(\ref{eq1}). 
% There are mainly two reasons: (1) The transverse waves propagate at the local Alfv\'{e}n speed inside and outside the flux tube, as seen in Figure \ref{Fig3}(A); while the waves propagate at the kink speed (give by Equation (\ref{eq1})) only near the tube boundary \citep{goossens2009}. (2) 
Because in our magnetic field configuration, Equation (\ref{eq1}) can only provide the phase speed as a function of $z$; however, here with a high spatial resolution, we are also interested in the horizontal distribution of phase speed and magnetic field. We note that such a choice may lead to some errors, especially within the flux tube region, which will be further discussed in Section \ref{sec:conclusion}. In Figure \ref{Fig4}(B), we compared the Alfv\'{e}n speed in the model and the wave propagation speed obtained from wave tracking. Despite some noticeable discrepancies, the overall correspondence between the two is reasonably good.

The calculated magnetic field is shown in Figure \ref{Fig3}(B), and its relative error (compared to the input value, $B_\text{los}$, which was calculated in the same manner as $\rho_\text{los}$) is presented in Figure \ref{Fig3}(C). The contours that correspond to the values of $\pm$30\% indicate that the relative error is less than 30\% for most pixels. Figure \ref{Fig4}(C) and (D) display histograms of $B_\text{seis}$ and its relative error. Statistically, there is a slight overestimation of the magnetic field by $\sim$5\% (with a standard deviation of 17\%). Given that the 5\% bias is quite small, the results support the reliability of the coronal seismology technique. Further discussion on the error distributions in Figure \ref{Fig3}(C) and the cause of the overestimation can be found in Section \ref{sec:conclusion}.

\subsection{After Degrading}\label{subsec:after}

\begin{figure}
   \centering
   \includegraphics[width=\textwidth, angle=0]{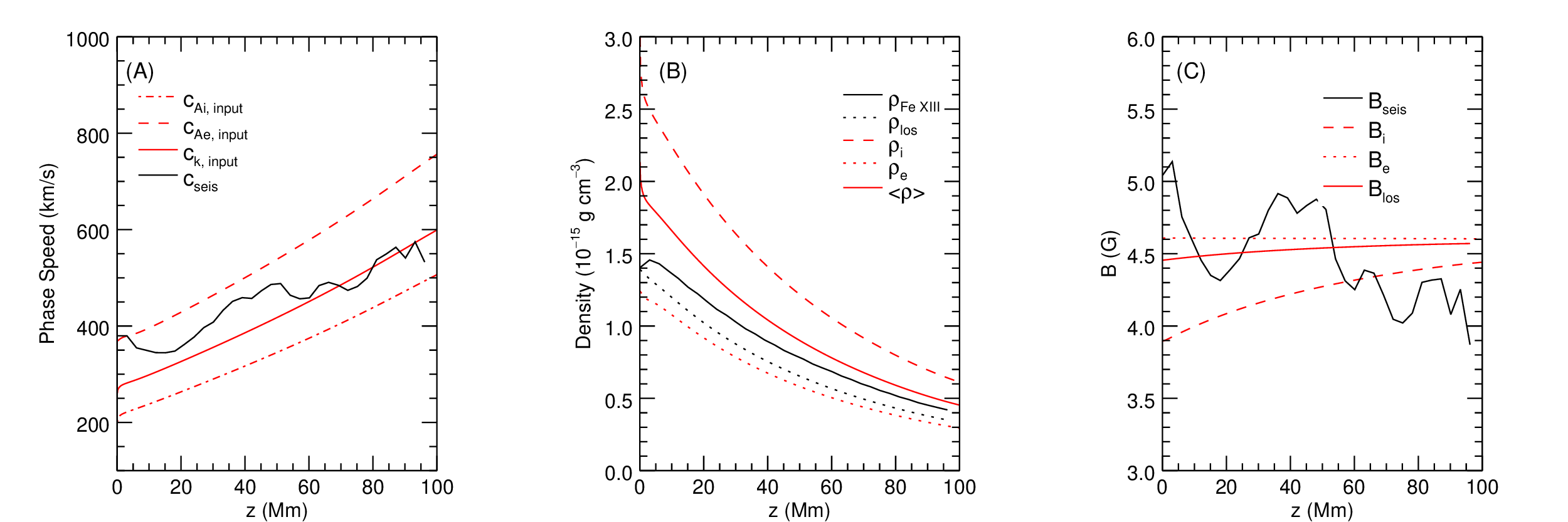}
   \caption{Phase speed, density, and magnetic field as a function of height ($z$). In panel (A), the red lines correspond to input values, including the internal Alfv\'{e}n speed (dot-dashed line), the external Alfv\'{e}n speed (dashed line), and the kink speed (solid line). The solid black line represents the wave propagation speed inferred from the wave tracking. In panel (B), the solid black line corresponds to the density derived from the Fe \textsc{xiii} intensity ratio ($\rho_\text{Fe XIII}$), and the dotted black line corresponds to the input value ($\rho_\text{los}$). The red lines represent the density in the model (similar to Figure \ref{Fig1}(D)), including the internal density $\rho_\text{i}$ (dashed line), the external density $\rho_\text{e}$ (dotted line), and the average density $\langle\rho\rangle=(\rho_\text{i}+\rho_\text{e})/2$ (solid line). In panel (C), the solid black line depicts the magnetic field inferred from coronal seismology ($B_\text{seis}$). The dashed and dotted red lines correspond to the internal and external magnetic fields ($B_\text{i}$ and $B_\text{e}$) in the model, respectively. The solid red line represents the input magnetic field ($B_\text{los}$).}
   \label{Fig6}
\end{figure}

We then degraded the spatial and temporal resolutions to match those of CoMP observations, with a pixel size of approximately 3.3 Mm and a cadence of 36 s. We focused on the central pixels (from $y = -1.65$ Mm to $y = 1.65$ Mm), which fully encompass the flux tube (diameter $\sim$2 Mm).
At this resolution, the tube boundary and finer structures are unresolved, similar to the case in observational studies \citep{yang2020ScChE, yang2020sci}.

We can now track the physical parameters along the tube axis. Figure \ref{Fig6} shows the phase speed, density, and magnetic field as a function of height $z$. In panel (A), the phase speed derived using the wave tracking method ($c_\text{seis}$) is compared to the characteristic speeds in the model (input values). The $c_\text{seis}$ closely matches the input kink speed ($c_\text{k, input}$), which lies between the internal and external Alfv\'{e}n speeds.

Panel (B) shows the density derived from the Fe \textsc{xiii} intensity ratio ($\rho_\text{Fe XIII}$), which also falls between the internal density ($\rho_\text{i}$) and external density ($\rho_\text{e}$) in the model. The $\rho_\text{Fe XIII}$ is slightly higher than the emissivity-weighted density along the LOS ($\rho_\text{los}$), similar to the pre-degradation results shown in Figure \ref{Fig4}(A).  When calculating the magnetic field with Equation (\ref{eq2}), the density value should ideally be the average of internal and external density (i.e., $\langle\rho\rangle=(\rho_\text{i}+\rho_\text{e})/2$), because Equation (\ref{eq2}) is derived from Equation (\ref{eq1}) when assuming $B_\text{i}=B_\text{e}$. However, in practice, $\rho_\text{Fe XIII}$ is used, which is slightly lower than $\langle\rho\rangle$. It means that using $\rho_\text{Fe XIII}$ as a representative of $\langle\rho\rangle$ can lead to a slight underestimation. The density underestimation is about 12-20\% based on Figure \ref{Fig6}(B), and the resulting error in $B_\text{seis}$ would be minimal (less than 5\%) since the density is taken the square root of.

In fact, the deviation of $\rho_\text{Fe XIII}$ from $\langle\rho\rangle$ can vary with the filling factor, which represents the fraction of the pixel occupied by the flux tube. Given the current pixel size (3.3 Mm), LOS integration length (8 Mm), and flux tube radius (1 Mm), the filling factor is approximately 12\%. If the filling factor is lower, the $\rho_\text{Fe XIII}$ will be closer to $\rho_\text{e}$, increasing the density underestimation. The maximum underestimation depends on the density contrast $\zeta=\rho_\text{i}/\rho_\text{e}$, with the deviation factor given by:
\begin{equation}
    \alpha=\frac{\rho_\text{Fe XIII}}{\langle\rho\rangle}\sim \frac{\rho_\text{e}}{(\rho_\text{e}+\rho_\text{i})/2}=\frac{2}{1+\zeta}\,.
\end{equation}
In our simulation, $\zeta$ was initially set to be 3, but after relaxation, it decreased to around 2 (see Figure \ref{Fig1}(D)). Such a value is comparable with previous observational estimates \citep{tian2012,verwichte2013,morton2021}. We tested the case with a reduced filling factor (~5\%), where a density contrast of 2 led to an underestimation of $\sim$30\%. Therefore, parameters like the filling factor and density contrast can be crucial when assessing the accuracy of coronal magnetic field measurements.

Figure \ref{Fig6}(C) presents the magnetic field inferred from coronal seismology ($B_\text{seis}$), compared to the internal ($B_\text{i}$), external ($B_\text{e}$), and emissivity-weighted ($B_\text{los}$) magnetic fields. The $B_\text{seis}$ ranges from 3.9 to 5.1 G, with errors below 15\%, indicating that the technique of coronal seismolgy can be used for reliable measurements of the coronal magnetic field strengths.

\section{Discussion and Conclusion}\label{sec:conclusion}

In this study, we tested the accuracy of the previously developed seismological techniques based on observations of propagating kink waves for deriving coronal magnetic fields. The results indicate that seismology-based measurements of the magnetic field are accurate and reliable to a large extent. 

In the high-resolution case, we obtained a 2D distribution of the magnetic field ($B_\text{seis}$) and its relative error. For most regions, the relative error is less than 30\%, as shown in Figure \ref{Fig3}(C). Statistically, the average magnetic field derived from coronal seismology is about 5\% larger than the input value. Although this case has a much higher resolution than the CoMP and UCoMP observations, the pixel size (62.5 km in the $z$ direction and 200 km in the $y$ direction) and cadence (12 s) can be comparable to those of the Cryogenic Near-Infrared Spectro-Polarimeter (Cryo-NIRSP; \citealp{Fehlmann2023}) and the Diffraction-Limited Near-Infrared Spectropolarimeter (DL-NIRSP; \citealp{Jaeggli2022}) of the Daniel K. Inouye Solar Telescope (DKIST; \citealp{Rimmele2020}). The intruments offer high-resolution coronal spectroscopic observations with the Fe XIII lines \citep{Schad2023}, and recently \cite{schad2024} successfully obtained a coronal LOS magnetogram with Cryo-NIRSP observation based on the Zeeman effect. With rapid repeated raster scans, it is possible that DKIST could also detect propagating transverse waves via Doppler velocity measurements, enabling seismological diagnostics of the plane-of-sky (POS) coronal magnetic field. In this way, DKIST observation can also provide us with maps of the POS magnetic field, but with a much higher spatial resolution compared to CoMP and UCoMP. Our results, particularly Figure \ref{Fig3}, can serve as a reference for such diagnostics. Combined with the Stokes-V measurements, DKIST may then be able to achieve measurements of the full magnetic field vector.

Nevertheless, we note that some errors appear in the seismologically inferred magnetic field. First, large errors appeared near the upper and lower boundaries (around $z=0$ and $z=100$ Mm). This is due to the Fourier filtering method used to subtract downward propagating wave components. As shown in Figure \ref{Fig2}(B), the upward propagating components of the Doppler velocity show some artificial velocity amplification near the upper and lower boundaries, especially after 250 s. This could affect the phase speed determination through wave tracking, leading to errors in $B_\text{seis}$ near the $z$ boundaries. Thus, in observations, it may be useful to exclude boundary signals before calculating phase speeds. 
Another contributing factor comes from the wave-tracking process itself. When calculating the phase speed, a certain number of data points along the propagating direction is required. Near the lower and upper boundaries, fewer data points will be available to perform cross-correlation, leading to larger uncertainties in the calculated phase speed and consequently in the inferred magnetic field.

Second, overestimations were observed inside or along the lateral boundaries of the flux tube. This is likely because we treat the phase velocity as the Alfv\'{e}n speed, which applies well to the regions away from the flux tube or cases with spatial averaging (see \citealt{yang2020sci} and Section \ref{subsec:after}). However, in this case, we modeled kink waves, and the flux tube can be well resolved. Thus, at the flux tube region, particularly the tube boundary, it would be more appropriate to apply Equation (\ref{eq1}) since the waves have a dominant kink wave characteristic \citep[e.g.,][]{goossens2009}. Calculating the magnetic field with $B_\text{seis}=c_\text{ph}\sqrt{\mu_0\rho}$ leads to overestimation at high-density regions, explaining the positive errors within the flux tube in Figure \ref{Fig3}(C). 
In future DKIST observations, we would suggest first applying Equation (\ref{eq2}) to diagnose the magnetic field outside the flux tube ($B_\text{e}$), then calculate the magnetic field inside the flux tube ($B_\text{i}$) with Equation (\ref{eq1}).

In addition, there are some other confusing patterns in Figure \ref{Fig3}(B) and (C). For instance, $B_\text{seis}$ and the relative error manifest fluctuations along the $z$ direction. It might be related to longitudinal oscillations excited by kink waves due to some non-linear effects \citep[e.g.,][]{Goldstein1978,DelZanna2001,terradas2004}. Further investigations are needed to understand these patterns.

When we degraded the spatial and temporal resolutions to approximately match those of CoMP, we found that the distributions of phase speed, density, and $B_\text{seis}$ along $z$ all show remarkable similarities to the input values (Figure \ref{Fig6}). Again we can notice a deviation near the upper and lower boundaries, particularly the lower one. However, within the height range of 20–80 Mm, the magnetic field error is generally less than 10\%. Additionally, the CoMP instrument has recently been upgraded to UCoMP, which has a slightly higher spatial resolution and a larger field of view. Our main conclusions should also apply to the case with the UCoMP observations \citep{yang2024}. We also tested the case when degrading to UCoMP’s pixel size ($\sim$2.2 Mm), and the results are largely similar to those shown in Figure \ref{Fig6}, with a slightly smaller error.

Another factor that may impact the phase speed and magnetic field measurements is the magnitude of the input magnetic field and phase speed. A stronger magnetic field and higher phase speed result in steeper slopes in the time-distance maps of Doppler velocity, which can reduce the accuracy of the wave-tracking method, particularly when the spatial and temporal resolutions are degraded. We ran a separate simulation with a background magnetic field of $\sim$10 G, which gives phase speeds of 1-2 Mm s$^{-1}$. We found a systematic underestimation of the magnetic field by approximately 30\% in this case. Nevertheless, the observed coronal magnetic field around $1.05-1.5R_\odot$ is 1-4 G \citep{lin2004,Gopalswamy2012,Kumari2019,yang2020sci,zhong2023NatSR}, which is more comparable with the case discussed in Section \ref{sec:res}. In conclusion, the coronal seismology technique based on propagating kink waves can provide reliable magnetic field measurements according to our forward modeling.

We note that this study only focuses on open coronal structures, including plumes in coronal holes and fan loops at the boundaries of active regions \citep{morton2015,banerjee2021}. Specifically, in our simulation, the magnetic flux tube is perpendicular to the solar surface, with both gravity and magnetic field aligned along the tube axis. However, the propagating kink waves can be detected not only in these structures but also in closed-field regions \citep[e.g.,][]{tomczyk2007,yang2020ScChE,yang2020sci}.
Our magnetic configuration can roughly describe one leg of a large-scale closed coronal loop, where the curvature can be neglected. For smaller loops where curvature cannot be ignored, our model is no longer applicable due to differences in gravitational stratification. Nevertheless, for such loops, phase speed along the axis often shows minimal variation \cite{McIntosh2011,threlfall2013,zhong2023NatSR}, thus similar to the case in \cite{magyar2018}, where a model without any vertical gradient in phase speed was used. Therefore, this study complements previous research by highlighting the importance of phase speed variation along the flux tube in coronal seismology.

Finally, we would like to mention that our model has some limitations. The vertical gradient of the magnetic field and the magnetic expansion are not included. Additionally, we didn't consider the effect of internal flows along the flux tube, which are frequently reported and can affect the apparent wave propagation speed \citep[e.g.,][]{soler2011,morton2015}.
Moreover, in real observations, there are often multiple flux tubes overlapping along the LOS, introducing further complexities that may affect wave-tracking accuracy and magnetic field measurements. In Figure 5 of \cite{yang2024}, comparisons between seismological results and global coronal MHD models revealed some discrepancies, particularly at higher latitudes where open field lines may dominate. Future work that incorporates more realistic models could offer a more sophisticated evaluation of the magnetic field measurements through coronal seismology and help clarify the discrepancies in \cite{yang2024}.

% \subsection{Conclusion}

\begin{acknowledgements}
This work is supported by the National Natural Science Foundation of China grant 12425301 and the Strategic Priority Research Program of the Chinese Academy of Sciences (Grant No. XDB0560000).
M.G. acknowledges support from the National Natural Science Foundation of China (NSFC, 12203030).
T.V.D was supported by the C1 grant TRACEspace of Internal Funds KU Leuven and a Senior Research Project (G088021N) of the FWO Vlaanderen. Furthermore, TVD received financial support from the Flemish Government under the long-term structural Methusalem funding program, project SOUL: Stellar evolution in full glory, grant METH/24/012 at KU Leuven. The research that led to these results was subsidised by the Belgian Federal Science Policy Office through the contract B2/223/P1/CLOSE-UP. It is also part of the DynaSun project and has thus received funding under the Horizon Europe programme of the European Union under grant agreement (no. 101131534). Views and opinions expressed are however those of the author(s) only and do not necessarily reflect those of the European Union and therefore the European Union cannot be held responsible for them. K.K. acknowledges support by an FWO (Fonds voor Wetenschappelijk Onderzoek – Vlaanderen) postdoctoral fellowship (1273221N).

\end{acknowledgements}

% \appendix                  %%appendicial material is supported

\bibliographystyle{raa.bst}
\bibliography{bibtex.bib} % your references Yourfile.bib

\label{lastpage}

\end{document}